\documentclass[12pt]{iopart}
\usepackage{iopams}
\usepackage{graphicx}  
\usepackage{dcolumn}   
\usepackage{bm}        
\usepackage{amssymb}   
\usepackage[latin1]{inputenc}
\usepackage{amsfonts}
\expandafter\let\csname equation*\endcsname\relax
\expandafter\let\csname endequation*\endcsname\relax
\usepackage{amsmath}
\usepackage{amsthm}
\usepackage{xcolor}
\usepackage{fontenc}
\usepackage{textcomp}
\usepackage{epstopdf}
\usepackage{hyperref}

\begin{document}
\title{Percolation in semicontinuum geometries}

\author{Jasna C. K.}
\ead{jasnack@cusat.ac.in, jasnack00@gmail.com}
\address{Department of Physics, Cochin University of Science and Technology, Cochin 682022, India.}
\author{V. Krishnadev}
\ead{vinkrishna05@gmail.com}
\address{Department of Physics, Cochin University of Science and Technology, Cochin 682022, India.}
\author{V. Sasidevan}
\ead{sasidevan@cusat.ac.in, sasidevan@gmail.com}
\address{Department of Physics, Cochin University of Science and Technology, Cochin 682022, India.}

\date{\today}
\begin{abstract}
We study percolation problems of overlapping objects where the underlying geometry is such that in $D-$dimensions, a subset of the directions has a lattice structure, while the remaining directions have a continuum structure. The resulting semicontinuum problem describes the percolation of overlapping shapes in parallel layers or lanes with positional constraints for the placement of the objects along the discrete directions. Several semicontinuum percolation systems are analyzed like hypercuboids with a particular focus on 2D and 3D cases, disks, and parallelograms. Adapting the excluded volume arguments to the semicontinuum setting, we show that for the semicontinuum problem of hypercuboids, for fixed side-lengths of the hypercuboids along the directions in which a lattice structure is maintained, the percolation threshold is always independent of the side-lengths along the continuum directions. The result holds even when there is a distribution for the side-lengths along the continuum directions. Trends in the variation of the thresholds, as we vary the linear measure of the shapes along the continuum directions, are obtained for other semicontinuum models like disks and parallelograms in 2D. The results are compared with those of corresponding continuum and lattice models. For the 2D and 3D models considered, using Monte Carlo simulations, we verify the excluded volume predictions for the trends and numerical values of the percolation thresholds. Very good agreement is seen between the predicted numerical values and the simulation results. The semicontinuum setting also allows us to establish a connection between the percolation problem of overlapping shapes in 2D continuum and the triangular lattice. We also verify that the isotropy of the threshold for anisotropic shapes and standard percolation universality class is maintained in the semicontinuum setting. 
\end{abstract}

\maketitle

\noindent{\it Keywords\/}:
Lattice percolation, Continuum percolation, Semicontinuum model, Excluded volume theory.

\section{Introduction}
Over the past several decades, percolation models that exhibit geometric phase transitions have garnered significant attention due to their theoretical and practical importance across various fields~ \cite{nadiv,croccolo,sander,philcox,cong,dechenaux,pokhrel,saberi}. These models effectively capture the inherent randomness of geometric structures present in many systems and show non-trivial critical behavior~\cite{isichenko,adler,soltani,sahimi}. Traditionally, percolation models have been studied in both lattice and continuum frameworks~\cite{stauffer,christensen,meester}. Lattice models are relatively easier to analyze whereas the continuum models are thought to represent better the random geometries encountered in several physical scenarios~\cite{Lin,JLin,thovert,hunt}. 

In recent works, researchers have investigated the percolation of overlapping and extended objects or shapes on lattices \cite{koza,koza1,brzeski}. In these models, objects are said to be overlapping when they share sites. Two objects are deemed connected if they overlap or two of their constituent sites are adjacent to each other. These models serve as a bridge between lattice and continuum approaches to percolation phenomena and find applications in problems such as transport in porous media \cite{koponen,koponen1}. Investigations into the discrete-to-continuum transition of overlapping aligned squares and cubes on a lattice have been conducted in~\cite{koza,koza1}. A similar study for hyperspheres was carried out in~\cite{brzeski}. 

In recent research, we demonstrated that models involving extended, aligned, and overlapping objects on lattices exhibit unique properties compared to their continuum counterparts~\cite{jasna}. Notably, these models display threshold behavior that depends on the width of the objects, which contrasts with the aspect-ratio-independent behavior observed in corresponding continuum models. For example, it was found that for overlapping and aligned rectangles placed on a square lattice, the percolation threshold decreases with increasing length when the width is one, increases with length when the width is greater than two, and remains constant when the width is two~\cite{jasna}. 

A lattice percolation model that involves extended overlapping objects is expected to transition to its corresponding continuum model by taking the limit where the lattice spacing approaches zero in all directions while proportionally increasing the size of the objects~\cite{koza,koza1}. In Ref.~\cite{jasna}, while discussing the percolation of extended shapes on lattices,  we proposed the possibility of taking the continuum limit of such lattice models only along a subset of the possible directions. In this work, we explore this scenario in detail, where, in a percolation model, the lattice structure is retained in one or more directions, while a continuum structure is assumed along the remaining directions. We call the resulting percolation model a semicontinuum one. Percolation models of overlapping objects in such semicontinuum geometries and their critical thresholds are the focus of the present work.

In simple terms, the semicontinuum model we consider here can be described as a percolation model involving overlapping shapes in parallel layers or lanes (see Fig.~\ref{connectivity}). Within each layer, the shapes can freely overlap. However, their placement and connectivity across layers are constrained by the discrete geometry present in that direction. Such a layered structure bears some resemblance to many real-world formations. Situations one can think of include materials with a layered structure, including rock sediments~\cite{bai,xu} or many biological tissues exhibiting layered structures~\cite{bini,hong}. The random geometries found in such layered formations may be effectively captured by semicontinuum geometries. Beyond their potential suitability to describe random structures in such settings, these models occupy an intermediate space between lattice and continuum geometries.  As we discuss in this paper, they can exhibit several non-trivial properties not observed in purely lattice or continuum systems. Understanding the semicontinuum models can therefore provide deeper insights into the role of discrete and continuum geometries in determining the physical properties of systems and how these properties evolve as we transition from discrete to continuum spaces.  

In this paper, we investigate the percolation properties of semicontinuum systems composed of overlapping objects. Specifically, we examine hypercuboids in general $D$ dimensions, disks in 2D, and parallelograms in 2D. The specific cases of hypercuboids in physically significant dimensions of 2D and 3D, i.e, that of rectangles and cuboids are considered in more detail. Making use of the approximate analytical method of Excluded Volume argument \cite{balberg,balberg1} adapted to the semicontinuum setting, we show that for fixed side-lengths of the hypercuboids along the directions in which a lattice structure is maintained, the percolation thresholds are found to be independent of the side-lengths along the continuum directions. This remains true even if there is a distribution for the side-lengths along the continuum directions. This behavior in the semicontinuum model is notably different from what is seen in both lattice and continuum systems. For example in the 2D lattice model, the trend of the threshold as the length of the rectangle varies depends on its width (it may increase, decrease, or remain constant, as discussed in Ref.~\cite{jasna}), whereas in continuum models, the threshold is always a fixed constant. For the semicontinuum model of disks in 2D, the percolation threshold is found to be a monotonically decreasing function of radius in contrast to the radius-independent threshold in continuum ~\cite{meester,balram}. Examining the semicontinuum model of parallelograms in 2D allows us to establish a connection between the  percolation problem of overlapping shapes in 2D continuum space and the triangular lattice. 

We use Monte Carlo simulations to verify the predictions regarding the trends in the thresholds and their numerical values of different 2D and 3D models considered. A very good agreement is obtained between the simulation results and the excluded volume predictions showing the usefulness of the latter in analysing the semicontinuum problems. Our simulation results also show that the percolation thresholds in the semicontinuum setting are isotropic in nature even for anisotropic shapes just as in lattice and continuum systems and that the standard percolation universality class is maintained in the semicontinuum setting. 

The paper is structured as follows. In Sec.~\ref{sec1}, we precisely define the semicontinuum model of percolation. In Sec.~\ref{sec2}, using a version of the excluded volume argument adapted to the semicontinuum geometry, we obtain numerical estimates and trends in the thresholds for several overlapping shapes in 2D and 3D, and analyze the case of hypercuboids in higher dimensions. Predictions in 2D and 3D are compared with simulation results. The impact of discreteness in a subset of directions upon percolation threshold is discussed in detail. Finally, we conclude the paper in Sec.~\ref{sec3}.

\section{Model definition}
\label{sec1}
\begin{figure}[hbt!]
    \centering
    \includegraphics[scale=0.7]{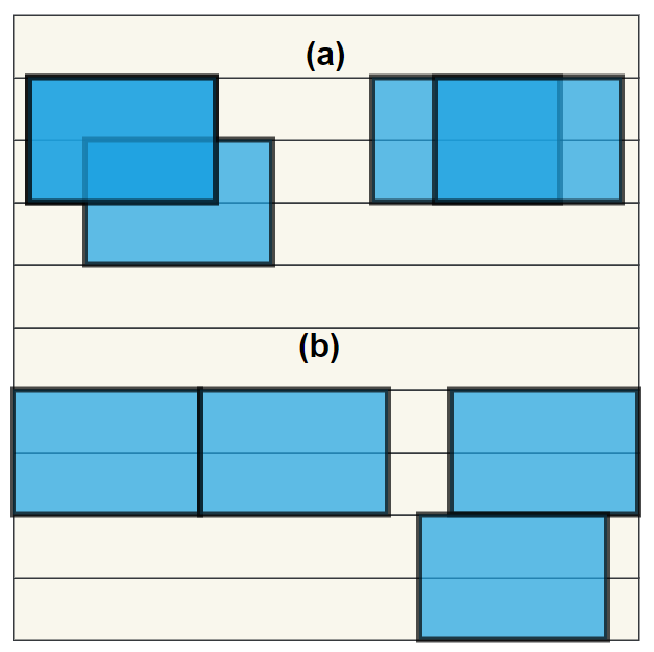}
   \caption{Placement and connectivity of rectangles in a $2D$ semicontinuum percolation model is shown with rectangles of length $3$ and width $2$ ($k_1 = 3, k_2 = 2$). Rectangles are placed randomly but with the constraint that width-wise they fit exactly $k_2$ lanes. They can freely overlap in the horizontal direction. Rectangles that overlap (Configurations (a)) or share an edge (Configurations (b)) either horizontally or vertically are considered connected.}
    \label{connectivity}
\end{figure}

We begin by defining the semicontinuum model in two dimensions, using the specific case of overlapping rectangles for clarity. Generalization to higher dimensions and other object shapes is easily done. In 2D, consider a set of $L$ parallel lanes or layers, each with length $L$ and width of one unit forming an $L\times L$ system. The width of the lane serves as our basic unit of measurement. Rectangles with length $k_1$ and width $k_2$ are randomly placed on these lanes so that, in width, they precisely cover $k_2$ lanes. Thus, $k_1$ can be any real positive number, while $k_2$ must be an integer multiple of the lane width. The shapes can overlap, but must adhere to the above constraint. Two rectangles are considered to be connected if they overlap or share an edge (see Fig.~\ref{connectivity}). We can see from the figure that a discrete geometry is maintained in the $Y$ (vertical) direction whereas a continuum geometry is maintained in the $X$ (horizontal) direction. This lane or layered geometry can be viewed as resulting from taking a square lattice of size $L\times L$ and letting the lattice spacing go to zero in only one direction (horizontal or $X$ direction in Fig.~\ref{connectivity}).

The impact of the discrete geometry in the vertical direction on the connectivity of shapes is evident when we consider width-one ($k_2 = 1$) rectangles. For these rectangles, overlapping is not possible vertically, but they can still connect by sharing an edge in the $Y$ direction. For rectangles with greater widths, overlapping in the vertical direction is possible, but with the constraint that width-wise the objects must fit exactly into an integer number of lanes. Consequently, while the geometrical centers of the objects can have any real value for their $X$-coordinates, their $Y$-coordinates are restricted to discrete values. If we allow the lane width to approach zero while proportionately increasing the width of the rectangles, the model converges to the corresponding continuum percolation problem. Thus, the model lies between the lattice and continuum percolation models, and we refer to it as the semicontinuum model of percolation.

As we increase the number of rectangles, the system undergoes a percolation transition.  In a finite system, a spanning cluster appears for the first time at the critical point, marking the percolation transition. We can characterize the critical point or the percolation threshold in terms of the areal density $\eta$ or the covered area fraction $\phi$ as we do in the usual continuum percolation problems involving overlapping objects \cite{mertens,xia,baker}. The areal density $\eta$ refers to the average total area of all objects placed in a unit area of the plane, while the covered area fraction $\phi$ represents the average fraction of the plane occupied by the objects. They are related by \cite{mertens},
\begin{equation}
    \phi=1-\exp(-\eta)
\label{eq0}
\end{equation}

\begin{figure}[hbt!]
    \centering
    \includegraphics[scale=0.60]{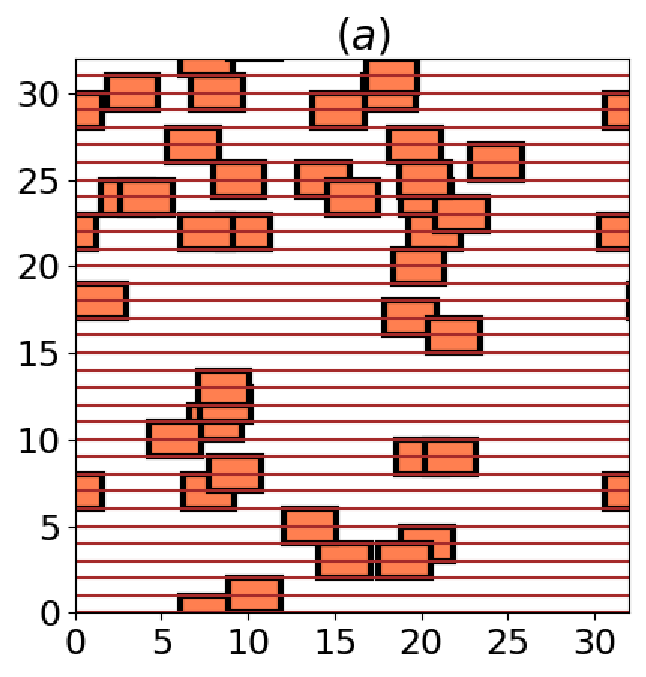}
     \includegraphics[scale=0.60]{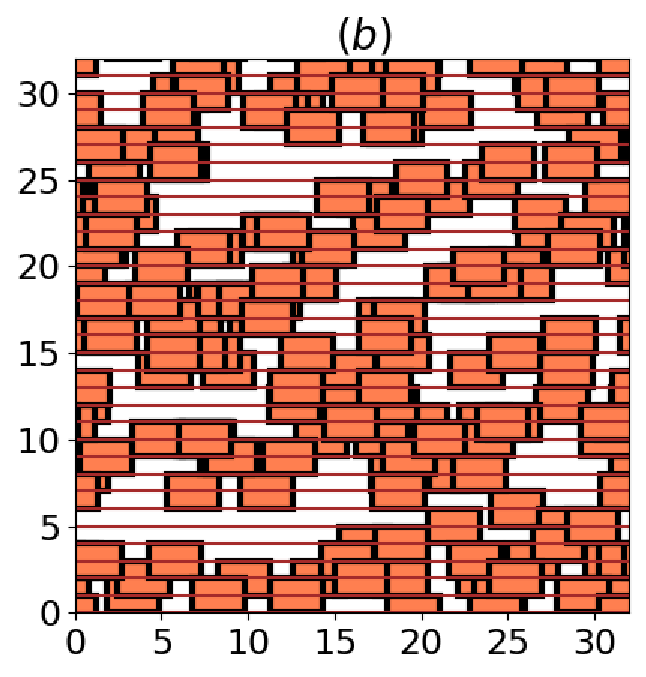}
    \caption{Sample configurations of a $2D$ semicontinuum system of overlapping rectangles of size $3 \times 2$ showing (a) System below the percolation threshold and (b) System above the percolation threshold with a spanning cluster.}
    \label{snapshots1}
\end{figure}

We define a spanning cluster as a connected path of objects that spans the system. In finite systems, we will consider spanning in the horizontal and vertical directions separately, allowing us to define horizontal and vertical percolation thresholds. However, as we will see, in the thermodynamic limit, both thresholds converge to a single value, similar to what is observed in lattice and continuum percolation problems involving anisotropic objects~\cite{jasna,klatt}. To minimize finite-size effects, periodic boundary conditions are assumed in both the directions. Note that the \textquotedblleft spanning\textquotedblright\; criteria with periodic boundary condition is the same as considering an arbitrary cut in the periodic system and checking for spanning. This has been used in some of the earlier works (See for example \cite{tarasevich}) and is expected to give the same threshold values in the infinite system limit as other criterion that can be used to define an effective threshold for finite systems. Typical configurations showing non-percolating and percolating phases in the 2D problem defined are shown in Fig.~\ref{snapshots1}

\begin{figure}[hbt!]
    \centering
    \includegraphics[scale=0.40]{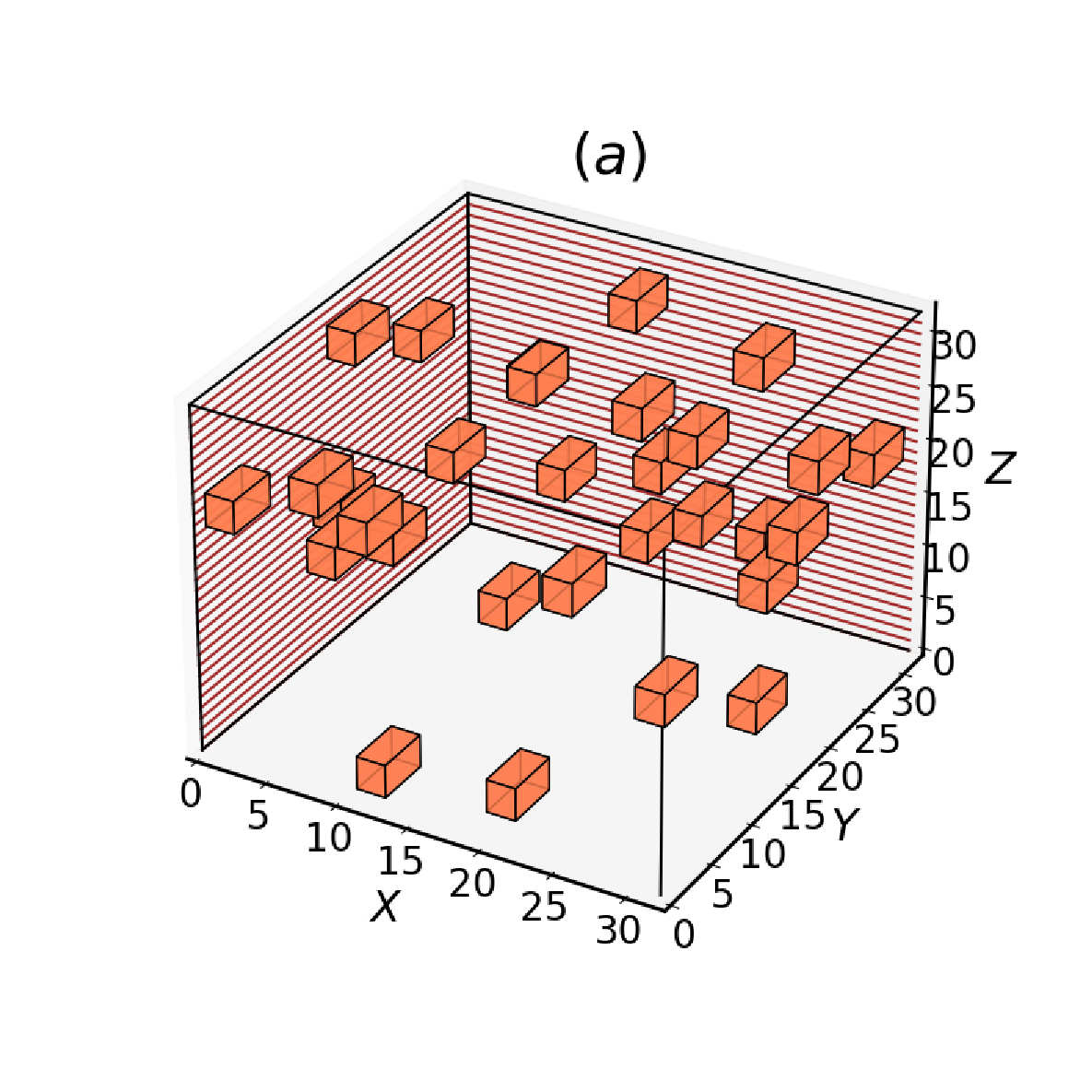}
     \includegraphics[scale=0.60]{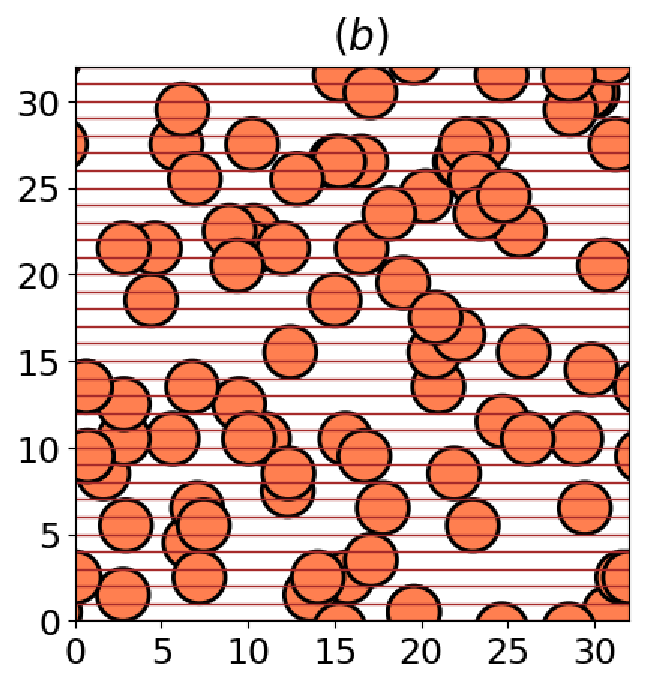}
    \caption{(a) A subcritical $3D$ semicontinuum system of overlapping cuboids of size $2 \times 4 \times 3$.  $X$ and $Y$ directions have continuum structures and $Z$ direction has a lattice structure. This implies that each cuboid fits exactly $3$ lanes in the $Z$ direction. There is no restriction on placement along $X$ and $Y$ directions. (b) A subcritical $2D$ semicontinuum system of overlapping disks of diameter $k=3$. $X$ direction has continuum structure and $Y$ direction has lattice structure. Each disk fits exactly $3$ lanes in the $Y$ direction. There is no restriction on placement along the $X$ direction.}
    \label{snapshots2}
\end{figure}

The semicontinuum model described above can be easily generalized to higher dimensions and other shapes.
In three-dimensional semicontinuum models, we have the option of making either one or two directions discrete. For example, consider the case of overlapping cuboids in a 3D semicontinuum geometry where two directions (say the $X$ and $Y$ directions) maintain continuous geometry, while the $Z$ direction has a discrete structure. This means that if $k_1$, $k_2$ and $k_3$ are the edge-lengths of the cuboids along the $X$, $Y$, and $Z$ directions respectively; then $k_1$ and $k_2$ can take positive real values while $k_3$ is restricted to positive integer values. The cuboids are randomly placed in these lanes, fitting exactly into $k_3$ lanes in the $Z$- direction. As in 2D, two objects are considered connected if they overlap or share a face. A typical configuration of a 3D semicontinuum model of cuboids described above is shown in Fig.~\ref{snapshots2}~(a). 

\section{Results and Discussion}
\label{sec2}
\subsection{Excluded volume theory for the semicontinuum models}
\label{subsec1}
The excluded volume argument has traditionally been widely used to analyze continuum percolation systems~\cite{balberg,balberg1,alvarez,balberg2}.  Let $V_{ex}$ represent the volume of the region around an object within which if another object is placed, they will overlap. The fundamental tenet of the excluded volume argument is that at the critical point, the total excluded volume $B_c$ which is the product of the number density $n_c$ of objects and their excluded volume $V_{ex}$ remains constant for similar shapes in a given dimension~\cite{balberg,balberg1}.  I.e, in the relation,
\begin{equation}
    n_c V_{ex}=B_c
    \label{eq1}
\end{equation}
 
$B_c$ is expected to be a constant for continuum systems composed of shapes that are related to each other by scaling and have the same orientation~\cite{klatt} in a given dimension (for example, disks and aligned ellipses).  
Since the number density of objects is given by $n=\eta/V$, where $V$ is the volume of an object, using Eq.~(\ref{eq0}) and Eq.(\ref{eq1}), the critical covered volume fraction~(ccvf) is given by,
\begin{equation}
    \phi_c=1-exp\left(-B_c\frac{V}{V_{ex}}\right)
    \label{eq2}
\end{equation}
Recently, Koza et al. extended the excluded volume argument to study the percolation of extended shapes on lattices \cite{koza}. In that context, the excluded volume is represented by a connectedness factor which is defined as the number of ways in which two basic percolating units can connect on the lattice~\cite{koza}.  
\begin{figure*}
    \centering
    \includegraphics[scale=0.65]{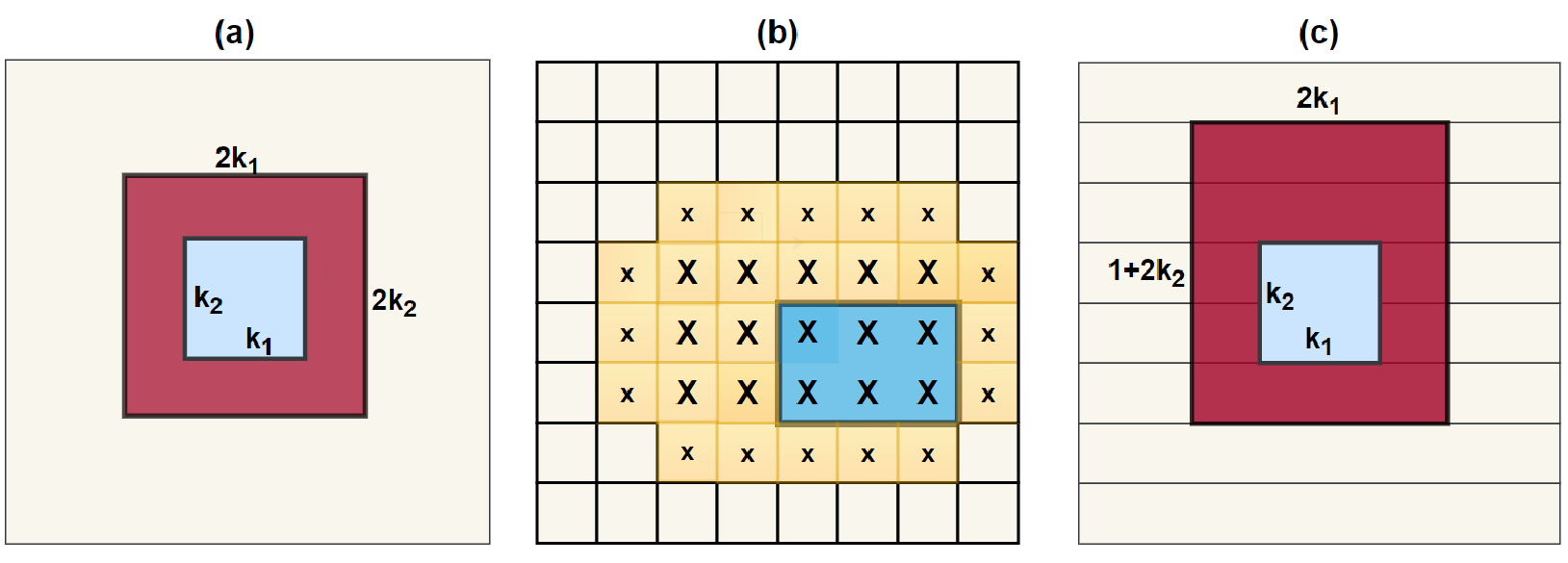}
    \caption{(a) Excluded area calculation for $2D$ continuum system of aligned, overlapping rectangles. For a rectangle of size $k_1\times k_2$,  a larger rectangle of size $2k_1 \times 2k_2$ forms the excluded area. Therefore, $V_{ex}= 2k_1 \times 2k_2$ (b) Excluded area calculation for $2D$ lattice model of aligned, overlapping rectangles. A rectangle of size $3 \times 2$ is shown for illustration. The cross-marked lattice sites or cells form the excluded area of the $3 \times 2$ rectangles in the figure.  The complete overlap of rectangles is not allowed. For a rectangle of size $k_1\times k_2$, a core region of size $(2 k_1 - 1) \times (2 k_2 - 1) $ (marked in larger cross) and a peripheral region of size $2 (2k_1-1) + 2 (2k_2-1)$ (marked in smaller cross) minus one form the excluded area. Therefore, $V_{ex}=(2k_1-1)(2k_2-1) + 2 (2k_1-1) + 2 (2k_2-1) - 1 = (2k_1+1)(2k_2+1) - 5$. (c) Excluded area calculation for $2D$ semicontinuum model of rectangles. For a rectangle of size $k_1\times k_2$,  a larger rectangle of size $2k_1 \times (2k_2 +1) $ forms the excluded area. Therefore, $V_{ex}= 2k_1 \times (2 k_2+1)$. Note how the discrete geometry along the vertical direction makes the excluded area calculation different from the continuum case.}
    \label{comparison_excludedarea}
\end{figure*}

We extend this formalism to semicontinuum problems by appropriately defining the excluded volume for such models. The key difference from continuum systems arises from the discrete nature of connectedness in a subset of directions. First, let us consider the 2D semicontinuum system of overlapping rectangles.  The difference in how the excluded volume $V_{ex}$ is calculated for continuum, lattice, and semicontinuum systems of overlapping rectangles of size $k_1 \times k_2$ is illustrated in~Fig.~\ref{comparison_excludedarea}. For the continuum system, $V_{ex} = 2k_1 \times 2 k_2$; for the lattice system $V_{ex} = (2k_1 + 1)(2k_2+1) - 5$; and for the semicontinuum system $V_{ex} = 2k_1 \times (2k_2 +1)$. As  $k_1$ and $k_2$ become larger, the excluded volumes for the lattice and semicontinuum models converge to that of the continuum model. We can draw some interesting conclusions for the semicontinuum system of rectangles by considering the ratio $V/V_{ex} = (k_1 k_2)/(2k_1(2k_2+1))$. The ccvf for the semicontinuum system of overlapping rectangles of size $k_1 \times k_2$ is given by,
\begin{equation}
    \phi_c^{k_1, k_2}=1-exp\left(-B_c\frac{k_2}{2(2k_2+1)}\right)
    \label{eq3}
\end{equation}
In fact, this is the expression for ccvf in the limit of $k_1 \rightarrow \infty$ of the lattice model of rectangles discussed in Ref.~\cite{jasna}. We can see from Eq.~(\ref{eq3}) that for the semicontinuum system, for a fixed width $k_2$, the percolation threshold $\phi_c^{k_1,k_2}$ is a constant, regardless of length $k_1$. For different widths, we get different constant thresholds which are independent of the length. This behavior contrasts sharply with the typical continuum scenario, where the percolation threshold is independent of the aspect ratio of the rectangles~\cite{klatt}. Let us consider the specific case of squares in a 2D square lattice ($k_1 = k_2 = k$) and compare it with the case of squares in a semicontinuum geometry. As $k \rightarrow \infty$, the threshold in both cases will approach the continuum value. From Eq.~(\ref{eq3}) with $k_2 = k$, we observe that in the semicontinuum geometry, $\phi_c^{k}$ increases monotonically with $k$.  This contrasts with the non-monotonic behavior of the threshold observed for overlapping squares on a lattice~\cite{koza}. 

\begin{figure}[hbtp]
    \centering
    \includegraphics[scale=0.70]{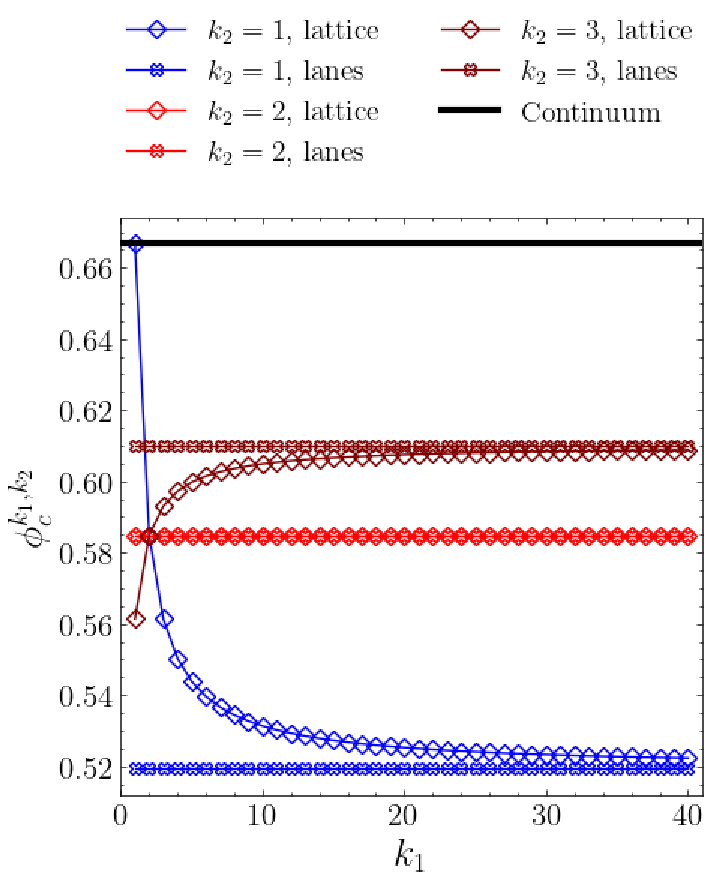}
    \caption{Comparison of excluded volume predictions of percolation thresholds for overlapping rectangles on a lattice and overlapping rectangles on lanes (semicontinuum model) for different widths $k_2$. The continuum threshold of rectangles is also shown ($\phi_c\approx 0.667$\cite{klatt}) for comparison.}
    \label{rect_phic_comparison}
\end{figure}

If we take a lattice model with rectangles of size $k_1\times k_2$, and take the continuum limit $k_1\rightarrow \infty$ keeping $k_2$ finite, we essentially obtain the semicontinuum model (See Ref.~\cite{jasna}). The approach of the lattice thresholds to the semicontinuum thresholds for different values of $k_2$ was considered in Ref.~\cite{jasna} and is depicted in Fig.~\ref{rect_phic_comparison}. The numerical value of $B_c$ is assumed to be the same as that of the corresponding continuum system ~\cite{koza}. We can see that, for widths $k_2=1$ and $k_2=3$, the thresholds in the lattice model approach the semicontinuum thresholds with opposing trends. However, for $k_2=2$, the lattice and semicontinuum thresholds coincide and are independent of $k_1$. The case of overlapping rectangles on purely lattice systems is discussed in detail in Ref.~\cite{jasna}.

Having described the 2D case in detail, let us consider the general case of hypercuboids in $D$-dimensional semicontinuum geometry. Let $k_i$ where $i=1,2,3,...D$ represent the linear measures of the hypercuboids along the $D$ dimensions respectively. Without loss of generality, we may assume that dimensions $1,2,3,,,d$ with $1 \leq d \leq D$ have continuum structure whereas the remaining dimensions $(d+1), (d+2)....,D$ have discrete or lattice structure. $d = D$ corresponds to the case of hypercuboids in a $D$- dimensional hypercubic lattice. The expression for the excluded volume for this hypercubic lattice system can be found by generalizing the 2D case in Fig.~\ref{comparison_excludedarea}~(b).
\begin{equation}
V_{ex} = \prod_{i = 1}^{D} (2 k_i - 1) + \left[2 \sum_{i = 1}^{D}\; \prod_{j=1,j\neq i}^{D} (2 k_j - 1)\right] - 1
\label{eq31}
\end{equation}
Here the first product term on the right-hand side gives the volume of the \textquoteleft core\textquoteright\; region of the excluded hypervolume of the hypercuboid and each term inside the summation gives the volume of the different \textquoteleft peripheral\textquoteright\; regions (See Fig.~\ref{comparison_excludedarea}~(b) for the definition of core and peripheral regions). We can rewrite Eq.~(\ref{eq31}) as,
\begin{equation}
    V_{ex} = \left[\prod_{i=1}^{D} (2k_i - 1) \right]\left[1+ 2 \sum_{i=1}^{D}\dfrac{1}{(2k_i-1)}\right] - 1
\label{eq32}
\end{equation}
The volume of the hypercuboids is given by $V = \prod_{i = 1}^{D}k_i$. Now the excluded volume for the semicontinuum system with continuum structure along $d$-dimensions can be obtained by taking the limit of $k_i \rightarrow \infty$ for $i = 1,2,3,...d$ in Eq.~(\ref{eq32}). Therefore, for the semicontinuum system of hypercuboids we get,
\begin{equation}
    V_{ex} = \left[\prod_{i=1}^{d} (2k_i) \right]\left[\prod_{i=(d+1)}^{D} (2k_i - 1) \right]\left[1 + 2 \sum_{i=(d+1)}^{D}\dfrac{1}{(2k_i-1)}\right]
\label{eq33}
\end{equation}
It is clear that the ratio $V/V_{ex}$ and hence the percolation threshold for the semicontinuum system of hypercuboids become independent of $k_i$ for $i = 1,2,3,...d$. In other words, for the semicontinuum percolation problem of hypercuboids, the percolation threshold is independent of the linear measures of the hypercuboids along the continuum directions. The 2D semicontinuum system of rectangles described earlier corresponds to $D=2$ and $d=1$.

As another specific case, consider a 3D semicontinuum model with cuboids having continuum geometry along the $X$ and $Y$ directions and a discrete geometry along the $Z$ direction (Fig.~\ref{snapshots2}~(a)). In this case, the expression for the ccvf becomes, 
\begin{equation}
\phi_c^{k_1,k_2,k_3}=1-\exp\left(-B_c \frac{k_3}{(8k_3+4)}\right)
\label{eq8}
\end{equation}

Eq.~(\ref{eq8}) tells us that the percolation threshold is a constant for a particular value of $k_3$ and is independent of $k_1$ and $k_2$. If we have a continuum geometry only along the $X$ direction and discrete geometry along the remaining two directions, the expression for the ccvf becomes the following.

\begin{equation}
    \phi_c^{k_1,k_2,k_3}=1-\exp\left(-B_c \frac{k_2 k_3}{8k_2k_3+4k_2+4k_3-6}\right)
    \label{eq9}
\end{equation}

From Eq.~(\ref{eq9}), we can conclude that for fixed values of $k_2$ and $k_3$, the percolation threshold is independent of $k_1$. In fact such independence of thresholds on the side-length(s) or measure(s) of the shapes along the continuum direction(s) in the model is expected because uniformly stretching the system along the continuum direction(s) maintains its connectivity structure. The same fact is reflected in the cancellation of factors corresponding to the continuum direction(s) in the ratio of volume to excluded volume of the objects in the excluded volume arguments leading to Eq.~(\ref{eq3}), Eq.~(\ref{eq8}), and Eq.~(\ref{eq9}).

It is easy to show that the independence of the threshold on linear measures of hypercuboids along the continuum directions holds even if we have a distribution for the linear measures along the $d$ continuum directions. For convenience, let us assume that the side lengths of the hypercuboid are drawn from a discrete distribution. Let $k_i^{j}$ with $j=1,2,....,M$ represent the $M$ possible values of the side-length $k_i$ for $i=1,2,...,d$. Let $p_i^{j}$ represent the probability that $k_i$ takes the value $k_i^{j}$. Then the (average) volume of the hypercuboid is,  
\begin{equation}
V =   \left[\sum_{j_1,j_2,...j_d = 1}^{M}  p_{1}^{j_1} p_{2}^{j_2}...p_{d}^{j_d}k_1^{j_1} k_2^{j_2}...k_d^{j_d}\right] \left[\prod_{i=d+1}^{D} k_i\right]
\end{equation}

and the average excluded volume of the hypercuboid is,
\begin{align}
V_{ex} = \left[\sum_{\substack{j_1,j_2,...j_d = 1\\ l_1,l_2,...l_d = 1}}^{M} p_{1}^{j_1} p_{2}^{j_2}...p_{d}^{j_d} p_{1}^{l_1} p_{2}^{l_2}...p_{d}^{l_d} \prod_{i=1}^{d} \left(k_i^{j_i} + k_i^{l_i}\right)\right]\times \left[\prod_{i=d+1}^{D}(2k_i-1)\right] \left[1+ 2 \sum_{i=d+1}^{D} \dfrac{1}{(2k_i - 1)}\right]
\end{align}
Expanding the product term inside the first square bracket on the right hand side gives $2^d$ terms each consisting of $d$ factors of $k_{i}^{j}$s. We can then carry out the  sum in the first square bracket over $d$ variables making use of the condition,
\begin{equation}
    \sum_{j=1}^{M}p_i^j = 1
\end{equation}
which is true for each $i$. This will give,
\begin{align}
    V_{ex} = 2^d \left[\sum_{j_1,j_2,...j_d = 1}^{M}  p_{1}^{j_1} p_{2}^{j_2}...p_{d}^{j_d}k_1^{j_1} k_2^{j_2}...k_d^{j_d}\right]
    \times \left[\prod_{i=d+1}^{D}(2k_i-1)\right] \left[1+ 2 \sum_{i=d+1}^{D} \dfrac{1}{(2k_i - 1)}\right]
\end{align}

It is clear that in the ratio $V/V_{ex}$, the factors involving the continuum dimensions will still cancel out, and the threshold remains independent of the distribution of $k_i$. The threshold is given by,
\begin{align}
    \phi_{c}^{k_{d+1},k_{d+2},...,k_D} = 1 - \exp\left(-B_c \frac{\left[\displaystyle\prod_{i=d+1}^{D} k_i\right]}{2^d\left[\displaystyle\prod_{i=d+1}^{D}(2k_i-1)\right] \left[1+ 2 \displaystyle\sum_{i=d+1}^{D} \dfrac{1}{(2k_i - 1)}\right]} \right)
\end{align}
with the appropriate value of $B_c$ for a given $D$.

\begin{figure}
    \centering
    \includegraphics[scale=0.90]{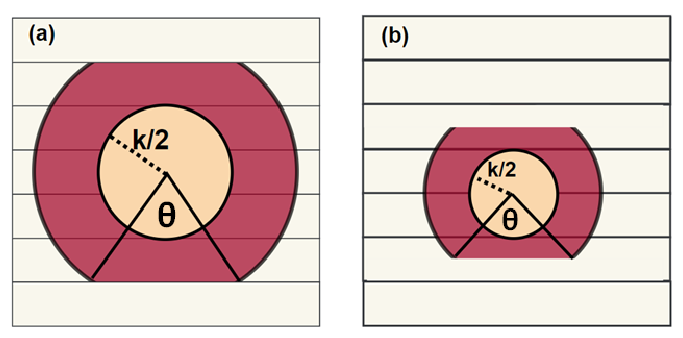}
    \caption{Excluded area calculation for disks on lanes. (a) For odd values of diameter $k$ ($k=3$ is shown), the centre of the disk is assumed to lie midway between two successive lanes and (b) for even $k$ ($k=2$ is shown), the centre is assumed to be on the lane itself. In both figures, the darker region marked around the disk together with the disk forms the excluded area. Note how the vertical length of the excluded area is restricted to cover exactly an integer number of lanes in (a) and (b). $\theta = 2\arccos\left(\frac{k-1/2}{k}\right)$ is marked in the diagram. $V_{ex}$ is calculated as the area of a circular disk of radii $k$ minus the area of the two missing minor segments on top and bottom. The excluded area is less than the continuum excluded area by a factor of $k^2\left(\theta-\sin(\theta)\right)$.}
    \label{vex_disk}
\end{figure}

We may extend the formalism to the semicontinuum problems involving other shapes as well. Here we consider the case of disks in $2D$. For a disk of diameter $k$, where $k$ is an integer multiple of lane-width and $k \geq 2$, the excluded area in the semicontinuum problem takes the form shown in Fig.~\ref{vex_disk}. Note that the diameter of the disks should be greater than a single lane width to have a percolation transition in the model. For odd and even values of $k$, we may assume that the centers of the disks occupy the midpoints between two successive lanes and the points on the lanes respectively. From Fig.~\ref{vex_disk}, it is easy to see that the excluded volume in both cases is given by,
\begin{equation}
    V_{ex}=k^2\left(\pi-\theta+\sin(\theta)\right)
    \label{eq_disk}
\end{equation}
Where $\theta = 2\arccos\left(\frac{k-1/2}{k}\right)$. Hence the expression for ccvf $\phi_c^k$ is given by,
\begin{equation}
    \phi_c^k=1-\exp\left(-B_c\frac{\pi}{4\left(\pi-\theta+\sin(\theta)\right)}\right)
    \label{eq_disk1}
\end{equation}
with the appropriate value of $B_c$. As $k$ increases, the percolation threshold $\phi_c^k$ monotonically decreases and approaches the corresponding continuum value $1-\exp\left(-B_c/4\right)$ as $k\rightarrow \infty$.

In the following section, we present the simulation results for the semicontinuum models of rectangles and disks in 2D, and cuboids in 3D and verify the above results. Additionally, we discuss the case of parallelograms in lanes and its relation to the percolation of overlapping parallelograms on a 2D triangular lattice~\cite{jasna}.

\subsection{Simulation Results}
\label{subsec2}
\subsubsection{Squares and Rectangles in 2D:}
\label{subsubsec1}

\begin{figure}[hbtp]
\centering
\includegraphics[scale=0.64]{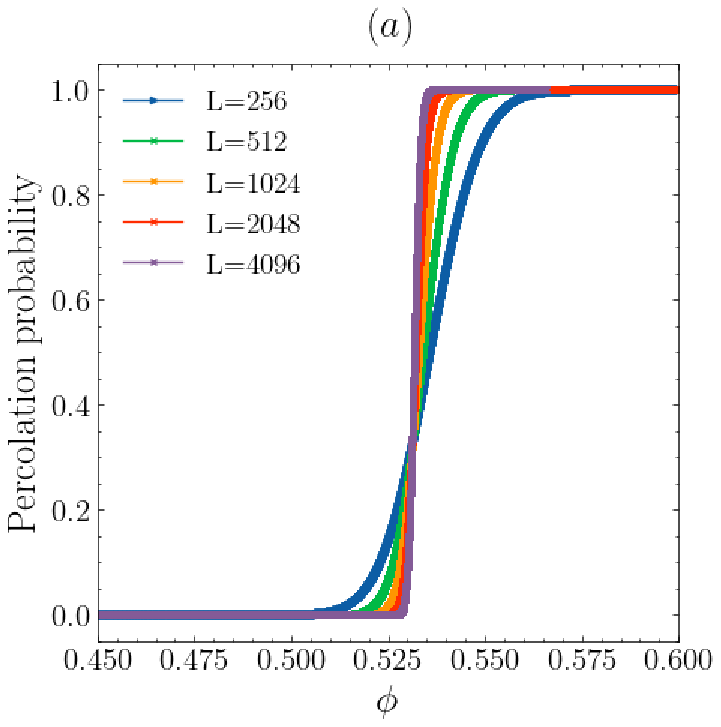}
\includegraphics[scale=0.64]{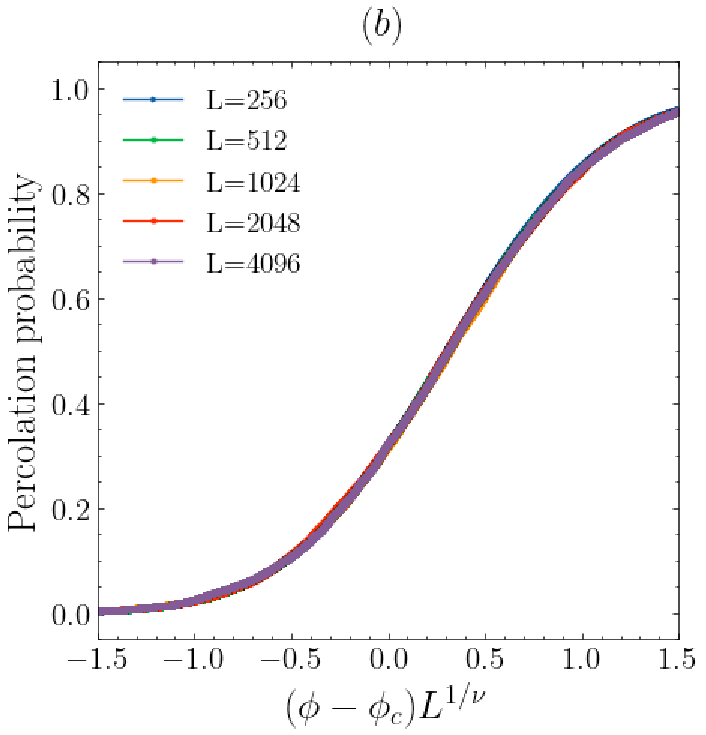}
\caption{(a) Percolation probability vs covered area fraction $\phi$ for rectangles in lanes with $k_1=2$, $k_2=1$. Vertical percolation probability is plotted. (b) Data collapse for the semicontinuum system of $2\times 1$ rectangles. Percolation probability is plotted against $(\phi-\phi_c)L^{1/\nu}$ for different $L$ with $\phi_c \approx 0.5311$ and $1/\nu =0.75$. Curves for different $L$ fall on top of each other.}
\label{percolation_prob}
\end{figure}

We first consider the $2D$ semicontinuum percolation problem of overlapping squares and rectangles. The objects are randomly placed in an $L \times L$ system one by one until a spanning cluster appears for the first time in vertical and horizontal directions. The process is repeated multiple times to determine the horizontal and vertical percolation probabilities as functions of the covered volume fraction. This is done by determining the fraction of realizations in which the respective type of spanning cluster appears for the first time. We use the Newman-Ziff algorithm for determining the thresholds~\cite{newman} utilizing a grid-based union-find algorithm for the semicontinuum geometry which is similar to the algorithm used in Ref.~\cite{mertens} for continuum systems. We divide the $L\times L$ semicontinuum system into several grids of size $k_1 \times k_2$. We ensure that $L$ is always an integer multiple of $k_1$ in the $X$ direction and a multiple of $k_2$ in the $Y$ direction which simplifies the process of checking for overlapping. System sizes considered range from a minimum of $L = 32$ to a maximum of $L=4096$ ($L=8190$ for larger $k$ squares) and the number of realizations considered ranges from a minimum of $10^4$ to a maximum of $10^6$. Typical plots of percolation probability against covered area fraction $\phi$ obtained for a few different system sizes are shown in Fig.~\ref{percolation_prob}~(a). 

From the percolation probability curves, we evaluate the effective percolation threshold $\phi_c(L)$ for different finite system sizes $L$, as well as the width of the transition region $\Delta(L)$, by fitting the curves with an error function $\frac{1+erf[(\phi-\phi_c(L))/\Delta(L)]}{2}$~\cite{rintoul}. To calculate the actual threshold in the thermodynamic limit $\phi_c(\infty)$, we plot $\phi_c(L)$ against $\Delta(L)$ which obey the scaling relation~\cite{stauffer},
\begin{equation}
    \phi_c(L) = \phi_c(\infty) + \textrm{Const.}\times\; \Delta(L)
    \label{eq10}
\end{equation}
The  $Y$-intercept of the straight line fit gives the true threshold $\phi_c(\infty)$. We calculate the horizontal and vertical thresholds separately. Typical plots are shown in Fig.~\ref{phieff_2x1}~(a). For rectangles, the horizontal and vertical effective thresholds for finite system sizes will be different as a result of the anisotropy of the shape. However, as we can see from Fig.~\ref{phieff_2x1}~(a), both converge to a single threshold value in the infinite system limit. This threshold isotropy is similar to what is seen for the percolation of extended objects in lattices and continuum systems~\cite{jasna, klatt}.

The percolation threshold values obtained from the simulations for squares of different sizes are listed in Table.~\ref{threshold_table} along with the excluded volume result based on Eq.~(\ref{eq3}). To obtain the numerical values of the thresholds from Eq.~(\ref{eq3}), we have assumed that the value of $B_c$ is the same as that of the corresponding continuum system ($B_c \approx 4.3953711$~\cite{koza}). The theoretical and simulation results are shown in Fig.~\ref{threshold_plot} in which a comparison with the results of the corresponding model on lattice system from \cite{koza} is also shown.
We can see that the threshold values increase monotonically for the semicontinuum model and there is a very good agreement between the theoretical predictions and simulation results. ({Note that the non-monotonicity observed for overlapping squares on a lattice in~\cite{koza} is only due to a single point. However, the comparison of the case of overlapping cubes on a lattice~\cite{koza} with the corresponding semicontinuum model also shows the same contrasting behavior between lattice and semicontinuum models (See Sec.~\ref{subsubsec5} below)). The excellent agreement seen even for relatively small values of $k$ validates the assumption that the critical connectivity of rectangles is the same (i.e. value of $B_c$) for continuum and semicontinuum models of overlapping squares. We see similar good agreement between simulation results and excluded volume predictions for other shapes like disks and rectangles indicating the validity of this assumption in general.

\begin{figure}[hbt!]
    \centering
    \includegraphics[scale=0.64]{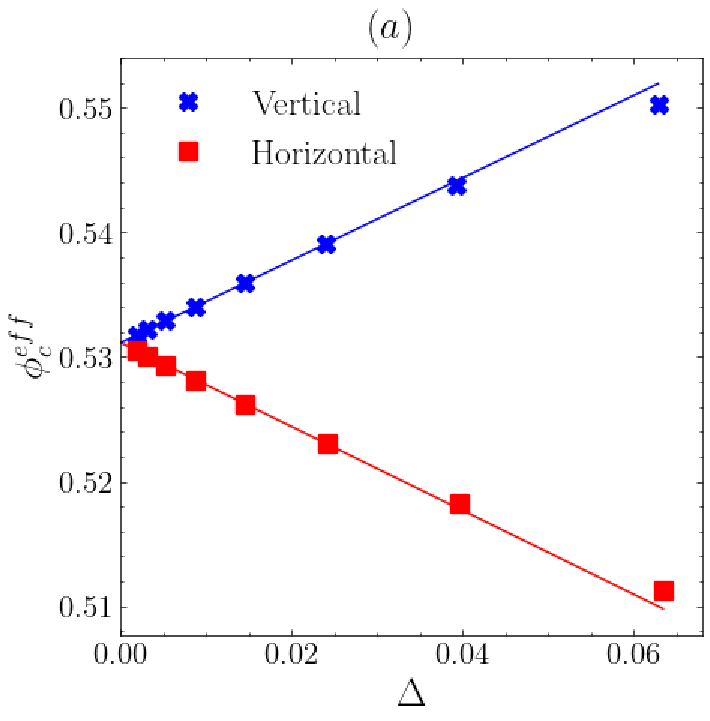}
    \includegraphics[scale=0.64]{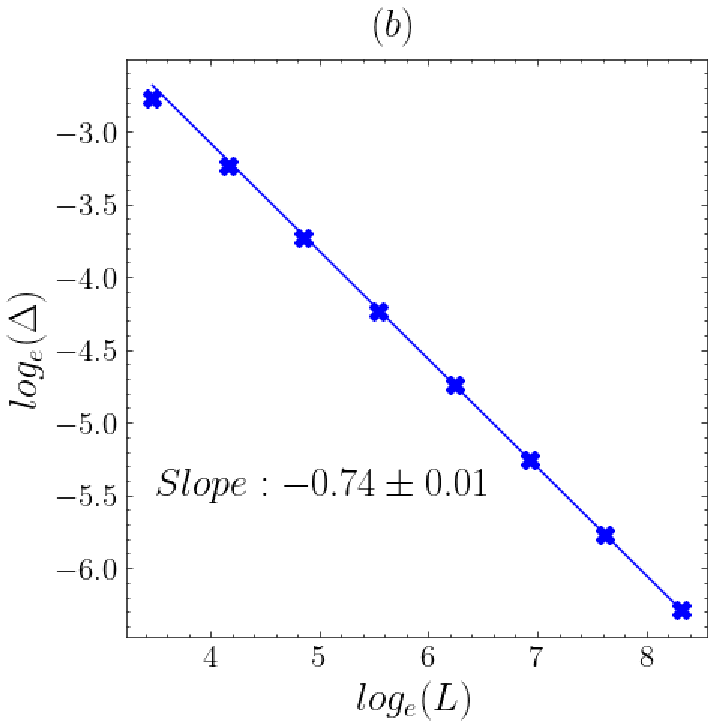}
    \caption{ (a) $\phi_c^{eff}$ vs width $\Delta$ for horizontal and vertical percolation for $2\times 1$ rectangles in lanes. The best straight-line fits with the largest five system sizes are shown. The $y$- intercepts give the true thresholds in the thermodynamic limit. Here for horizontal percolation, the $y$-intercept is $\phi_c^{k_1,k_2} = 0.53112(1)$ and for vertical percolation it is $\phi_c^{k_1,k_2}=0.53114(2)$. (b) $\log - \log$ plot of the width of the percolation curves $\Delta$ with system size $L$ along with the best straight line fit for $2 \times 1$ rectangles. The straight-line fit is obtained with the largest five system sizes. Slope of the straight line gives $1/\nu  = 0.74\pm0.01$ (See Eq.~(\ref{eq11})).}
    \label{phieff_2x1}
\end{figure}

\begin{table}[hbt!]
\caption{\label{threshold_table} Percolation threshold $\phi_c^k$ for various values of $k$ for overlapping squares in lanes determined from simulations and Eq.~(\ref{eq3}). Percolation threshold obtained from simulation is the average of horizontal and vertical thresholds (see text).}
\begin{indented}
\item[]\begin{tabular}{ccc}
\br
k & $\phi_c^k$ Average & $\phi_c^k$ theory\\
\mr
1 & 0.53111(3)& 0.51932\\
2 & 0.58909(3)& 0.58483\\
3 & 0.61230(3)& 0.61010\\
4 & 0.62476(4) & 0.62347\\
5 & 0.63269(3) & 0.63173\\
6 & 0.63807(5)& 0.63735\\
7 & 0.64184(8) &0.64141\\
8 & 0.64477(5)& 0.64449\\
9 & 0.64721(3)& 0.64690\\
10 & 0.64909(7)& 0.64884\\
11 & 0.65061(8) & 0.65044\\
12 & 0.65200(4) &0.65177\\
13 & 0.65308(2)&0.65290\\
14 & 0.65416(8) & 0.65387\\
15 & 0.65482(3)& 0.65471\\
20 & 0.65785(4) & 0.65769\\
\br
\end{tabular}
\end{indented}
\end{table} 

\begin{figure}[hbtp]
    \centering
    \includegraphics[scale=0.75]{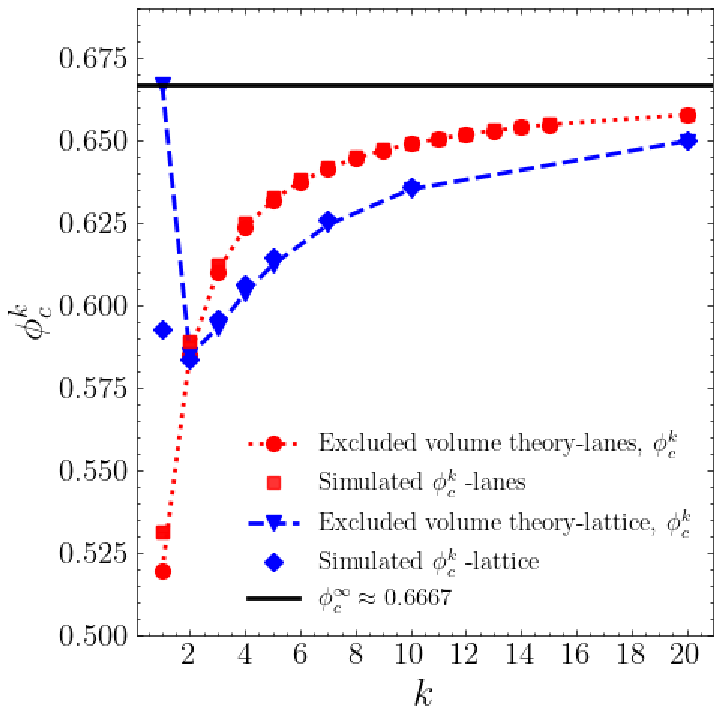}
    \caption{Variation of percolation threshold $\phi_c^k$ with $k$ for overlapping squares in lanes compared with that for overlapping squares on a square lattice given in~\cite{koza}. Simulation results and excluded volume results are shown in both the cases. Corresponding continuum value is also shown for comparison ($\phi_c^{\infty}\approx 0.6667\cite{mertens})$. Semicontinuum model of squares shows a monotonic increase of threshold with $k$ whereas lattice models show a non-monotonic behavior (See Ref.~\cite{koza}). Note that the non-monotonicity observed for overlapping squares on a lattice is only due to a single data point. However, this contrasting behavior between the lattice and semicontinuum models is also seen in the 3D case (Sec.~\ref{subsubsec5}). }
    \label{threshold_plot}
\end{figure}

Percolation thresholds for rectangles with different $k_1$ and $k_2$ values are obtained in a similar way.  Results for widths $k_2=1, 2$ and $3$ are listed in Table.~\ref{threshold_table_rect}. Again, the simulation results are in agreement with the predictions of the excluded volume theory. We can also see that the threshold is nearly a constant for a particular width confirming its length independence. 

\begin{table*}[hbt!]
\caption{\label{threshold_table_rect} Percolation threshold $\phi_c^{k_1,k_2}$ for rectangles in lanes for widths $k_2 = 1, 2$ and $3$ and for increasing values of length $k_1$. Results from simulations and Eq.~(\ref{eq3}) are given. Percolation threshold obtained from simulation is the average of horizontal and vertical thresholds.}
\begin{indented}
\item[]\begin{tabular}{cccccc}
\br
$k_1$ & $k_2$ & $\phi_c^{k_1,k_2}$ -vertical & $\phi_c^{k_1,k_2}$ -horizontal & Average $\phi_c^{k_1,k_2}$ & $\phi_c^{k_1,k_2}$ theory \\
\mr
1 & & 0.53122(5) & 0.53099(4) & 0.53111(3)&  \\
2 & & 0.53114(2) & 0.53112(1) & 0.53113(1)& \\
3 & 1 & 0.53111(2) & 0.53114(3)& 0.53113(2) & 0.51932\\
4& & 053110(3) & 0.53112(2)& 0.53111(2) & \\
5 & & 0.53111(4) &0.53119(2)& 0.53115(2) & \\
\hline
1 &  & 0.58912(3)& 0.58904(2) & 0.58908(2) & \\
2 & & 0.58909(4)& 0.58909(3)  & 0.58909(3)& \\
3 & 2 & 0.58909(2) & 0.58906(2) & 0.58908(1)&0.58483\\
4& & 0.58909(7) & 0.58911(1)& 0.58910(4) & \\
5 & & 0.58904(2) &0.58918(9)& 0.58911(5) & \\
\hline
1 & & 0.61216(6)  &0.61224(4)  & 0.61220(4) &\\
2 & & 0.61233(2)  &0.61224(5) & 0.61229(3) & \\
3 & 3 & 0.61231(2) & 0.61229(6) & 0.61230(3) &0.61010\\
4& & 0.61229(3) &0.61227(2) & 0.61228(2) & \\
5 & & 0.61228(2) &0.61231(3)& 0.61230(2) & \\
\br
\end{tabular}
\end{indented}
\end{table*}

We also evaluate the correlation length exponent $\nu$ for the semicontinuum model of rectangles. We use the scaling relation~\cite{stauffer},
\begin{equation}
    \Delta(L) \propto L^{-1/\nu}
\label{eq11}
\end{equation}
The slope of the $\log - \log$ plot between $\Delta(L)$ and $L$ gives the exponent $1/\nu$. A sample plot for the semicontinuum problem of $2\times 1$ rectangles is shown in Fig.~\ref{phieff_2x1}~(b). The obtained values for all systems involving rectangles of different sizes are found to be in line with the standard value of $1/\nu=3/4$ for 2D site percolation~\cite{stauffer,christensen}. As a confirmation, we verify that a good scaling collapse is obtained for the percolation probability $P(\phi)$ with the determined values of the threshold and the correlation length exponent. $P(\phi)$ is expected to follow the scaling relation \cite{stauffer},
\begin{equation}
    P(\phi)= f((\phi-\phi_{c})L^{\frac{1}{\nu}})
    \label{eq12}
\end{equation}
Hence plots of $P(\phi)$ with $(\phi-\phi_c)L^{1/\nu}$ for different values of $L$ should show a good scaling collapse with the correct values of the threshold $\phi_c$ and the exponent $\nu$. The data collapse obtained for different $k_1$ and $k_2$ confirms that the semicontinuum models of rectangles belong to the standard percolation universality class. A sample data collapse curve is shown in Fig.~\ref{percolation_prob}~(b).

\subsubsection{Other 2D models: Disks and parallelograms in lanes}
\label{subsubsec4}
In this section, we consider two other semicontinuum systems in 2D. The first one is a 2D semicontinuum model of disks of single radius described in Sec.\ref{subsec1}. Disks of diameter $k$ are placed randomly in lanes until a spanning cluster appears (see Fig.~\ref{snapshots2}~(b)).  The threshold values are obtained in a similar manner described in Sec.\ref{subsubsec1}. The percolation thresholds obtained for the diameter values of $k=2,3,4,5,6,7$ are listed in Table~\ref{disc_table}. We can see a decreasing trend in the threshold with increasing $k$ as predicted by Eq.~(\ref{eq_disk1}). Again, excellent agreement is seen between the predicted values and the simulation results.

\begin{table}[hbt!]
\caption{\label{disc_table} Percolation threshold $\phi_c^{k}$ for semicontinuum disk models of different diameter values $k$ determined from simulations and Eq.(\ref{eq_disk1}). Percolation threshold obtained from simulation is the
average of horizontal and vertical thresholds. Threshold calculated from Eq.(\ref{eq_disk1}) uses $B_c \approx 4.512348$ (taken from corresponding continuum model) \cite{xun}.}
\begin{indented}
\item[]\begin{tabular}{ccc}
\br
$k$ & Average $\phi_c^{k}$ & Theory $\phi_c^{k}$\\
\hline
2 & 0.7341(5) & 0.7324\\
3 & 0.7059(5)& 0.7064\\
4 & 0.6944(3)& 0.6953\\
5 & 0.6903(3)&  0.6902\\
6 & 0.6866(5) & 0.6869\\
7 & 0.6848(4) & 0.6847\\
\br
\end{tabular}
\end{indented}
\end{table}

The second shape we consider is overlapping parallelograms. We can view this system as the semicontinuum limit of overlapping parallelograms on a triangular lattice (Fig.~\ref{TRlattice}). Here, we first consider a prediction made in Ref.~\cite{jasna} that on a triangular lattice, parallelograms of width $k_2 = 1$  will have a constant percolation threshold independent of its length. The numerical value of the threshold is given by $\phi_c=1-\exp(-B_c/6)$ where $B_c$ is expected to be the same as that of the continuum percolation problem of rectangles. Table.~\ref{TRlattice_table} 
 lists the threshold values for widths $k_2 = 1, 2$ and $3$. Note that $k_1 = k_2 = 1$ corresponds to the usual site percolation on the triangular lattice for which the threshold is known exactly to be $1/2$ ~\cite{sykes}.

From Table.~\ref{TRlattice_table}, we can see that for widths $k_2=2$ and $k_2=3$, there is a clear increasing trend in threshold and the values are in reasonable agreement with the predictions. However, an increasing trend in the threshold values is seen for width $k_2=1$ as well which shows a deviation from the prediction of a constant value. Thus the excluded volume theory prediction of a $k_1$-independent threshold doesn't quite hold up in this case although from Fig.~\ref{deviation}~(a), we can see that the percentage deviation of the thresholds from their asymptotic value is the lowest for width $k_2=1$ for any $k_1$. A similar comparison is shown for rectangles on square lattices in Fig.~\ref{deviation}~(b).  We can see that in this case, the percentage deviation of the threshold values from their asymptotic value is the smallest for width $k_2=2$ for any $k_1$. For the case of $k_2 = 2$, except for $k_1 = 1$, there is very little deviation (less than 1\%) seen in the
threshold values.

\begin{figure*}[hbt!]
    \centering
    \includegraphics[scale=0.66]{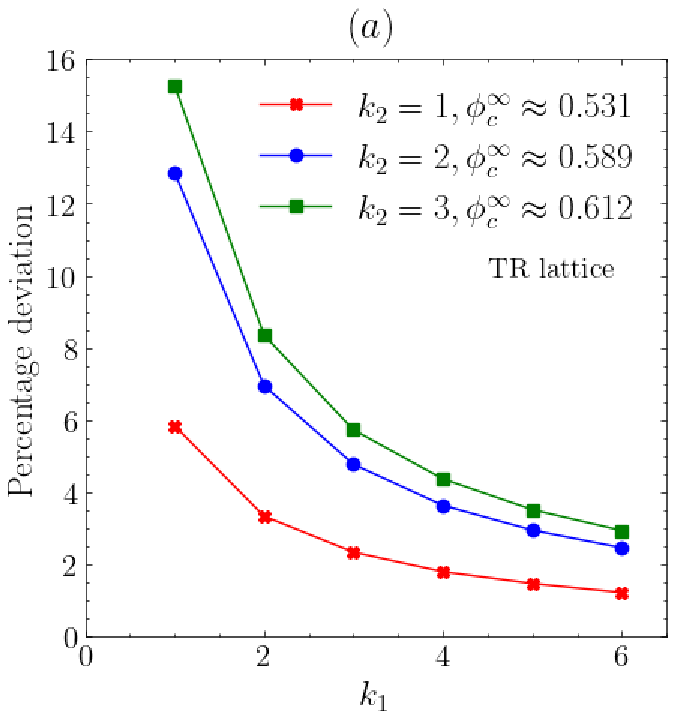}  
    \includegraphics[scale=0.66]{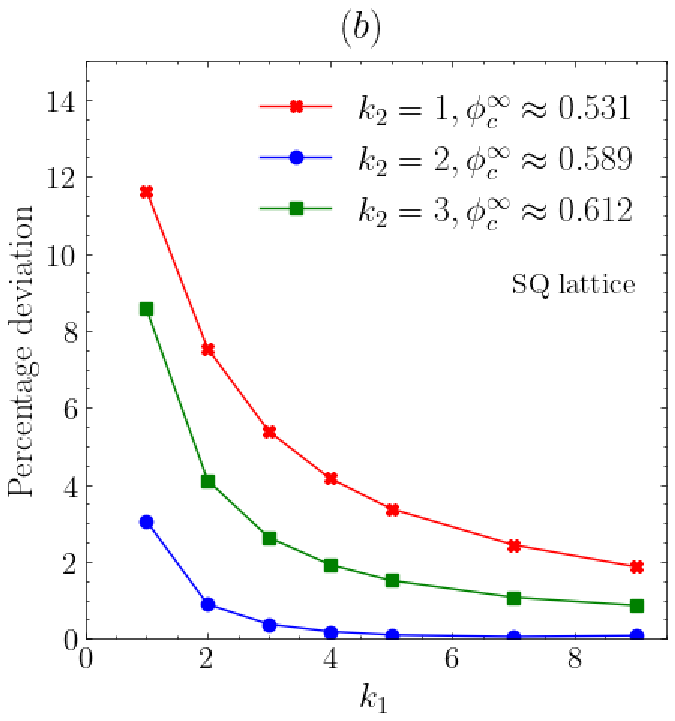}  
    \caption{(a) Percentage deviation of the thresholds obtained from simulation from their approximate asymptotic values ($\phi_c^\infty$) for different $k_1$ for parallelograms of widths $k_2=1$, $k_2=2$ and $k_2=3$ on a triangular lattice. (b) Percentage deviation of the thresholds obtained from simulation from their approximate asymptotic values ($\phi_c^\infty$) for different $k_1$ for rectangles of widths $k_2=1$, $k_2=2$ and $k_2=3$ on a square lattice. The simulation data is an improvement over the one given in Ref.~\cite{jasna}. The approximate asymptotic values ($\phi_c^\infty$) are taken from Column 5 of Table.~\ref{threshold_table_rect}.}
    \label{deviation}
\end{figure*}

Now, for the corresponding semicontinuum problem, we can imagine taking the triangular lattice and making the $X$ direction continuous while maintaining a discrete geometry in the $Y$ direction. We can see that with periodic boundary conditions applied, the semicontinuum problem of parallelograms of size $k_1 \times k_2$ (note that here $k_2$ refers to the number of lanes the object covers in the vertical direction and not the Euclidean length) is the same as that of rectangles of the same size. Thus we expect the same thresholds for the two systems which we can verify by comparing Tables~\ref{threshold_table_rect} and \ref{parallelogram_semi_table}. We also expect that parallelograms in a continuum geometry have the same threshold as squares or rectangles in a continuum as the shapes are related to each other by a shearing transformation~\cite{torquato,modenov}, which we verify in Table.~\ref{parallelogram_continuum_table}.

\begin{figure*}[hbt!]
    \centering
    \includegraphics[scale=0.64]{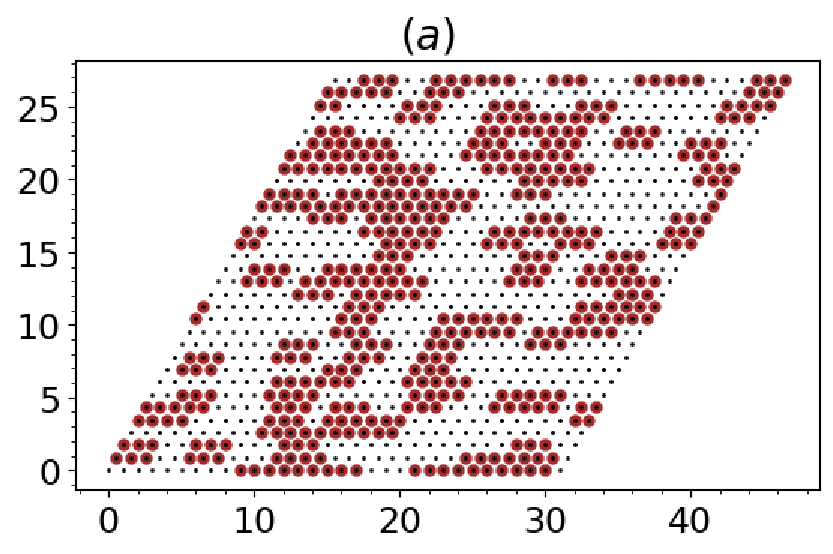}  
    \includegraphics[scale=0.51]{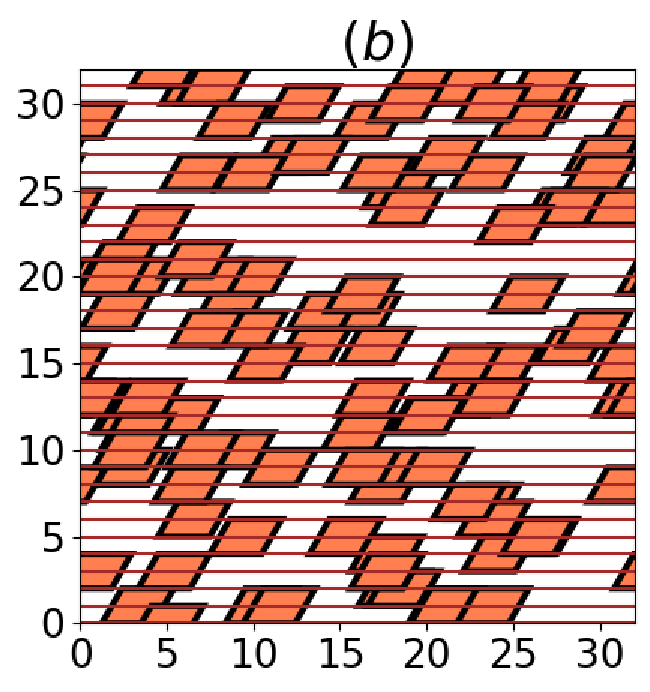}  
    \caption{(a) Aligned overlapping parallelograms of size $k_1=3$ and $k_2=2$ on a triangular lattice. (b) Semicontinuum model of overlapping parallelograms of size $k_1=3$ and $k_2=2$ on 2D lanes (Note that in the semicontinuum case, $k_2$ refers to the number of lanes the parallelogram covers in the vertical direction and not the Euclidean length). We get the semicontinuum lane structure in (b) by making the horizontal direction continuous in (a) while maintaining the lattice structure along the vertical direction.}
    \label{TRlattice}
\end{figure*}

\begin{table}[ht]
\caption{\label{TRlattice_table} Percolation threshold $\phi_c^{k_1,k_2}$ for parallelograms of width $k_2=1$, $k_2=2$ and width $k_2=3$ on triangular lattice from simulations and the excluded volume results~\cite{jasna}. Percolation threshold obtained from simulation is the average of horizontal and vertical thresholds.}
\begin{indented}
\item[]\begin{tabular}{cccc}
\br
$k_1$ & $k_2$& Average $\phi_c^{k_1,k_2}$ & $\phi_c^{k_1,k_2}$ Theory \\
\hline
1& & 0.49999(2) & \\
2& & 0.51328(3)& \\
3& 1 & 0.51858(3)& 0.51932\\
4& & 0.52145(5)& \\
5& & 0.52319(8)& \\
6& &0.52451(11)& \\
\hline
1& & 0.51330(3)& 0.51932\\
2&  &0.54805(2) & 0.55029\\
3& 2 & 0.56089(3) & 0.56138\\
4& & 0.56758(6) & 0.56708\\
5& &0.57163(6)& 0.57056\\
6& & 0.57450(13) & 0.57289\\
\hline
1& & 0.51856(2)& 0.51932\\
2& & 0.56089(5)& 0.56138\\
3& 3& 0.57695(5)& 0.57682\\
4& & 0.58526(4)& 0.58483\\
5& & 0.59047(4)& 0.58974\\
6& & 0.59402(10)& 0.59305\\
\br
\end{tabular}
\end{indented}
\end{table}

\begin{table}[ht]
\caption{\label{parallelogram_semi_table} Percolation threshold $\phi_c^{k_1,k_2}$ obtained from simulations of a semicontinuum system of parallelograms of width $k_2=1$ and $k_2=2$. The percolation threshold is the average of horizontal and vertical thresholds. Theoretical values are obtained using Eq.~(\ref{eq3}).}
\begin{indented}
\item[]\begin{tabular}{cccc}
\br
$k_1$ & $k_2$ & Average $\phi_c^{k_1,k_2}$ & Theory $\phi_c^{k_1,k_2}$ \\
\hline
1& & 0.53109(2)& \\
2&  &0.53116(2)& \\
3& 1 & 0.53114(3)& 0.51932\\
4& & 0.53114(3) & \\
5& & 0.53112(5) &\\
6& & 0.53114(3) & \\
\hline
1& & 0.57555(1)& \\
2& & 0.58909(5) & \\
3& 2 & 0.58909(2)& 0.58483\\
4& & 0.58908(3)& \\
5& & 0.58905(3)& \\
6& & 0.58911(4) & \\
\br
\end{tabular}
\end{indented}
\end{table}

\begin{table}[hbt!]
\caption{\label{parallelogram_continuum_table} Percolation threshold $\phi_c^{k_1,k_2}$ obtained from simulations of a continuum system of parallelograms of width $k_2=1$ and $k_2=2$. Percolation threshold is the average of horizontal and vertical thresholds.}
\begin{indented}
\item[]\begin{tabular}{ccc}
\br
$k_1$ & $k_2$ & Average $\phi_c^{k}$\\
\hline
5& 1 & 0.66662(5)\\
4& 1 & 0.66658(6)\\
\hline
5& 2 & 0.66679(4)\\
4& 2  & 0.66678(2)\\
\br
\end{tabular}
\end{indented}
\end{table}

\subsubsection{Cuboids in 3D :}
\label{subsubsec5}
As an example of a semicontinuum system in 3D, we consider overlapping cuboids in lanes having continuous $X$ and $Y$ coordinates while having discrete $Z$ coordinates (see Fig.\ref{snapshots2}~(a)). The thresholds are determined as in the 2D case and are given in Table.~\ref{3d_semi_table} along with the theoretical predictions based on Eq.~(\ref{eq8}). We can see that, in agreement with predictions of Eq.~(\ref{eq8}), threshold $\phi_c^{k_1,k_2,k_3}$ is independent of $k_1$ and $k_2$, for a given value of $k_3$. The numerical estimates of the thresholds are obtained by assuming $B_c \approx 2.5978$ ~\cite{koza} to be that of the corresponding continuum problem. 

From Table~\ref{3d_semi_table}, we can also infer that the percolation threshold for cubes in the semicontinuum shows an increasing trend with the side-length. Although we have only two data points in Table.~\ref{3d_semi_table} corresponding to cubes ($k_1=k_2=k_3 = 2$ \& 3), already this is in contrast to the case of cubes on a lattice whose threshold values first show a decrease and then an increase to go to its continuum value of $\approx 0.277$~\cite{koza}. 

\begin{table*}[hbt!]
\caption{\label{3d_semi_table} Percolation threshold $\phi_c^{k_1,k_2,k_3}$ for semicontinuum cuboid model with $k_3=2$ and $k_3=3$ determined from simulations and Eq.~(\ref{eq8}). Percolation threshold obtained from simulation is the average of the thresholds in three different directions.}
\begin{tabular}{cccccccc}
\br
$k_1$ & $k_2$ & $k_3$ & $\phi_c^{x}$ & $\phi_c^{y}$ & $\phi_c^{z}$ & $\phi_c$ Average &$\phi_c$ theory (See Eq.(\ref{eq8}), $B_c \approx 2.5978$~\cite{koza})\\
\mr
2 & 2 & & 0.2300(1)& 0.2307(2)& 0.2304(2)& 0.2304(1) &\\
2 & 3 & 2 & 0.2299(8)& 0.2294(1)& 0.2298(8)& 0.2297(4)&0.2288\\
3 & 3 & & 0.2302(2)& 0.2298(2)& 0.2307(1)& 0.2302(1)&\\
\mr
2 & 4 & & 0.2448(8)& 0.2436(10)& 0.2441(6)& 0.2442(5)&\\
2 & 2 & 3 & 0.2421(9)& 0.2437(3)& 0.2421(3) & 0.2426(3)&0.2429\\
3 & 3 & & 0.2440(4)& 0.2433(3) & 0.2439(4)& 0.2437(2)&\\
\br
\end{tabular}
\end{table*}

\section{Conclusion}
\label{sec3}
In this work, we studied percolation problems of overlapping objects in geometries that have lattice structures along certain directions and continuum structures along the remaining ones. In 2D and 3D, the resulting geometry corresponds to a lane/layered structure with objects distributed randomly in and across the layers obeying the constraints present due to the lattice structure in a subset of the directions. We show that the percolation problem of overlapping objects in such semicontinuum geometries have several interesting properties concerning the variation of percolation thresholds with the linear measures characterizing the basic percolation units. 

We analyze a number of percolation problems of overlapping objects in 2D and 3D. In 2D, we considered the percolation of overlapping rectangles, disks, and parallelograms, and in 3D, we considered cuboids. We adapted the excluded volume argument to a semicontinuum setting and studied the dependence of the percolation threshold on the linear measures characterizing the basic percolation units considered. We show that in 2D, for rectangles, the percolation threshold is independent of the length (linear measure along the continuum direction) of the rectangles and is a constant for a given width (linear measure along the discrete direction), even when we have a distribution of lengths. This contrasts with the aspect-ratio-independent thresholds seen in continuum models. Aligned overlapping squares in lanes show a monotonic increase in threshold with increasing size of the squares. This is in contrast to the non-monotonic behavior for the threshold observed for overlapping squares on lattices~\cite{koza}. We analyzed overlapping disks and parallelograms in 2D semicontinuum geometries. For disks, we find that the threshold decreases monotonically with their radius. 

We find similar results for 3D systems. For cuboids, we show that the percolation threshold is independent of its linear measures along the continuum directions for fixed measures along the discrete directions. The predictions regarding the trends in the thresholds and their numerical values for the 2D and 3D systems are validated by numerical simulations.  A very good agreement is seen between the two both in terms of the trend and in terms of the numerical values.

In general, we show that for a semicontinuum system of hypercuboids with $d$ continuum directions,  the percolation threshold is independent of its linear measures along the $d$ directions. This remains true even if the linear measures of the hypercuboid along the continuum directions have a distribution.

For obtaining the numerical values of the thresholds, we implicitly assumed that the critical connectivity pattern remains the same for continuum and the semicontinuum systems for percolation problems involving similar shapes distributed with the same orientation. The good numerical match observed between the theoretical predictions and the simulation results validates this assumption. In addition, our results on parallelograms in 2D semicontinuum geometry implicitly show that the critical connectivity is the same for similar shapes in continuum and triangular lattice establishing a connection between the two. 

Models in lattice and continuum geometries have been extensively studied in several fields within Science, Engineering, and Mathematics. In many of them, studying how the properties of a system change as we go from a lattice to continuum geometry is a question of great significance. The semicontinuum geometry introduced here describes a system that lies between the two extremes of lattice and continuum geometries. We studied the specific problem of the percolation of overlapping objects in such geometries, but it is obvious that many other models may be considered in such geometries and could be of interest in many fields opening up directions for further investigations. A particular example is that of models with hardcore exclusion that have been extensively studied in lattices and show properties such as jamming and order-disorder transitions~\cite{slutskii,kundu}. 

Finally, we note that numerical values of the thresholds obtained here are based on the assumption that the connectivity at criticality is the same for similar shapes in continuum and semicontinuum systems, which allowed us to use the known values of $B_c$ from continuum systems. Our theoretical predictions of the numerical values in the models discussed turns out to be very accurate in most cases, justifying the assumption. However, theoretical treatments dedicated to semicontinuum geometries may be required to treat other percolation problems in such settings.

In conclusion, we show that models that lie in semicontinuum geometries show some unique behavior not seen in their lattice and continuum counterparts. Our study brings about a deeper understanding of the role of discrete and continuum geometries in determining the physical properties of systems and how these properties evolve as we transition from discrete to continuum spaces.

\section*{Code availability} Sample source code for our project is available on the GitHub repository. \\ \url{https://github.com/jasnaCK/Semicontinuum-Percolation}

\ack
\noindent Authors acknowledge the use of the high-performance computing cluster established at Cochin University of Science and Technology (CUSAT) under the Rashtriya Uchchatar Shiksha Abhiyan (RUSA 2.0) scheme (No. CUSAT/PL(UGC).A1/2314/2023, No: T3A).

\end{document}